\documentclass[twocolumn,superscriptaddress]{revtex4-1}

\usepackage[paperwidth=210mm,paperheight=297mm,centering,hmargin=2cm,vmargin=2cm]{geometry}
\usepackage{braket}
\usepackage{amsmath}
\usepackage{siunitx}
\usepackage{ragged2e}
\usepackage{bbold}
\usepackage{graphicx}
\usepackage{times}

\makeatletter
\renewcommand\@biblabel[1]{#1.}
\makeatother

\usepackage[dvipsnames]{xcolor}
\usepackage{hyperref}
\hypersetup{
  colorlinks   = true, 
  urlcolor     = MidnightBlue, 
  linkcolor    = MidnightBlue, 
  citecolor   = Maroon 
}

\begin{document}

\title{Demonstration of fault-tolerant Steane quantum error correction}

\author{Lukas Postler}
    \affiliation{Institut f\"{u}r Experimentalphysik, Universit\"{a}t Innsbruck, Innsbruck, Austria}
\author{Friederike Butt}
    \affiliation{Institute for Quantum Information, RWTH Aachen University, Aachen, Germany}
    \affiliation{Institute for Theoretical Nanoelectronics (PGI-2), Forschungszentrum J\"{u}lich, J\"{u}lich, Germany}
\author{Ivan Pogorelov}
    \affiliation{Institut f\"{u}r Experimentalphysik, Universit\"{a}t Innsbruck, Innsbruck, Austria}
\author{Christian D. Marciniak}
    \affiliation{Institut f\"{u}r Experimentalphysik, Universit\"{a}t Innsbruck, Innsbruck, Austria}
\author{Sascha Heu{\ss}en}
    \affiliation{Institute for Quantum Information, RWTH Aachen University, Aachen, Germany}
    \affiliation{Institute for Theoretical Nanoelectronics (PGI-2), Forschungszentrum J\"{u}lich, J\"{u}lich, Germany}
\author{Rainer Blatt}
    \affiliation{Institut f\"{u}r Experimentalphysik, Universit\"{a}t Innsbruck, Innsbruck, Austria}
    \affiliation{Alpine Quantum Technologies GmbH, Innsbruck, Austria}
    \affiliation{Institut f\"{u}r Quantenoptik und Quanteninformation, \"{O}sterreichische Akademie der Wissenschaften, Innsbruck, Austria}
\author{Philipp Schindler}
    \affiliation{Institut f\"{u}r Experimentalphysik, Universit\"{a}t Innsbruck, Innsbruck, Austria}
\author{Manuel Rispler}
    \email[Correspondence email address: ]{rispler@physik.rwth-aachen.de}
    \affiliation{Institute for Quantum Information, RWTH Aachen University, Aachen, Germany}
    \affiliation{Institute for Theoretical Nanoelectronics (PGI-2), Forschungszentrum J\"{u}lich, J\"{u}lich, Germany}
\author{Markus M\"uller}
    \affiliation{Institute for Quantum Information, RWTH Aachen University, Aachen, Germany}
    \affiliation{Institute for Theoretical Nanoelectronics (PGI-2), Forschungszentrum J\"{u}lich, J\"{u}lich, Germany}
\author{Thomas Monz}
    \affiliation{Institut f\"{u}r Experimentalphysik, Universit\"{a}t Innsbruck, Innsbruck, Austria}
    \affiliation{Alpine Quantum Technologies GmbH, Innsbruck, Austria}

\begin{abstract}
Encoding information redundantly using quantum error-correcting (QEC) codes allows one to overcome the inherent sensitivity to noise in quantum computers to ultimately achieve large-scale quantum computation.
The Steane QEC method involves preparing an auxiliary logical qubit of the same QEC code used for the data register. The data and auxiliary registers are then coupled with a logical CNOT gate, enabling a measurement of the auxiliary register to reveal the error syndrome. This study presents the implementation of multiple rounds of fault-tolerant Steane QEC on a trapped-ion quantum computer. Various QEC codes are employed, and the results are compared to a previous experimental approach utilizing flag qubits. Our experimental findings show improved logical fidelities for Steane QEC. This establishes experimental Steane QEC as a competitive paradigm for fault-tolerant quantum computing.
\end{abstract}

\maketitle

\section{Introduction}
Quantum computing has the potential to outperform classical machines by exploiting superposition and entanglement. To achieve this goal of enhanced computational capabilities, it is crucial to safeguard quantum information, e.g. by encoding it into stabilizer codes that protect against environmental and operational noise. By repeatedly measuring the stabilizer generators, we can detect noise without disrupting the logical computational state. The error and its location within the register are mapped onto the results of the stabilizer measurements, also referred to as error syndrome. We must prevent the spread of harmful errors by following the principles of quantum fault tolerance to conduct quantum computation on the encoded level while maintaining the expected scaling of logical error rates with physical error rates. This requirement implies experimental challenges in fault-tolerant (FT) logical state preparation, FT logical gates, and FT error correction.

Recent progress in achieving error-corrected universal quantum computation has been made through the development of FT QEC components in leading hardware architectures. In superconducting systems, significant strides have been made towards operating Kitaev's surface code, resulting in an operation fidelity that exceeds the break-even point~\cite{google2021exponential, krinner2022realizing, zhao2022realization, google2023suppressing}. Additionally, FT magic state preparation has been demonstrated in a superconducting experiment with fidelity beyond break-even~\cite{gupta2023encoding}. Ion-trap experiments have demonstrated FT stabilizer readout~\cite{hilder2022fault}, FT control of single logical qubits~\cite{Egan2021}, and FT repetitive QEC cycles~\cite{ryan2021realization}, with subsequent efforts aimed at implementing universal FT logical gate sets~\cite{postler2022demonstration, ryan2022implementing}. Meanwhile, practical experimental benefits of fault tolerance have been demonstrated in error-detecting codes, such as FT non-Clifford gates on multiple logical qubits in both superconducting and trapped-ion devices~\cite{Menendez2023}, FT one-bit addition as a small logical algorithm on three logical qubits~\cite{Wang2023}, the realization of Grover search utilizing encoded qubits in a trapped-ion device~\cite{Pokharel2022}, and the very recent demonstration of a larger logical quantum processor with neutral atoms~\cite{Bluvstein2023}.

The backbone of the successful operation of a fault-tolerant quantum processor is an efficient implementation of QEC cycles. Steane QEC minimizes the coupling between data and auxiliary qubits and therefore also perturbations of the data register. Thus it is a promising candidate for the efficient extraction of error syndromes on scalable error-correcting codes.

\section{Fault-tolerant quantum error correction}

\begin{figure*}[t]
	\centering
	\includegraphics[width=0.67\linewidth]{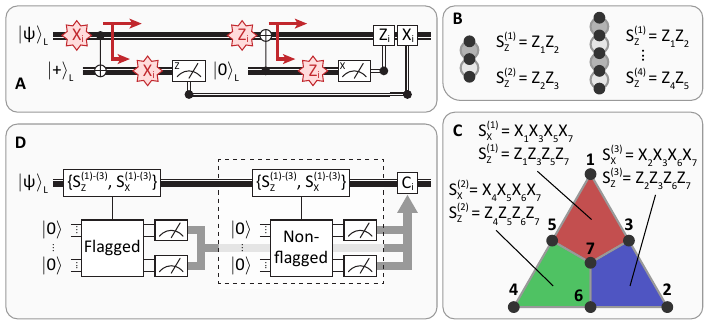}
	\caption{\justifying \textbf{Quantum error correction codes and methods for syndrome extraction.} (\textbf{A}) In Steane-type error correction, an auxiliary logical qubit is prepared in $|+\rangle_L$, then coupled to the initial logical qubit $|\psi\rangle_L$ with a transversal CNOT gate, and then measured in the computational basis. This procedure is repeated for an auxiliary logical qubit in $|0\rangle_L$ with an inverted transversal CNOT gate. A correction is applied to the corresponding data qubits based on the syndrome which is extracted from the measurements performed on the auxiliary logical qubit. (\textbf{B}) The repetition code encodes one logical qubit and its stabilizer generators are weight-$2$ Pauli-operators defined between neighboring qubits on a line. (\textbf{C}) The seven-qubit color code encodes one logical qubit into seven physical qubits. A code state is a $+1$ eigenstate of all six weight-$4$ stabilizer generators \{$S^{(i)}_X, S^{(i)}_Z$\} defined on the colored plaquettes. Pauli-$X$ (-$Z$) on all qubits corresponds to the logical operator $X_L$ ($Z_L$), which is up to multiplication with stabilizer generators equivalent to weight-$3$ operators. (\textbf{D}) Flag error correction includes the measurement of the stabilizer generators with flagged circuits. In case of non-trivial measurement outcomes of the flagged circuits, repeating the measurement of stabilizer generators is necessary. Different variants of flag-based QEC require different numbers of simultaneously available auxiliary qubits and remeasuring procedures. The flag QEC procedure used in this work is described in Appendix~\ref{app:seven_qubit_CC}.}
	\label{fig:overview_schemes_and_codes}
\end{figure*}

Given operations acting on logically encoded qubits, such as initialization, gate operations, and measurements, have to be constructed in a way that prevents dangerous propagation of errors. An error configuration is dangerous when an otherwise correctable number of errors spreads via entangling gates and turns into an error supported on a number qubits (referred to as the weight of the error) beyond the number of errors the code can correct. A circuit where this is precluded by design is called a fault-tolerant implementation and hence we refer to the corresponding operations as fault-tolerant.
In particular, this applies to the QEC block itself, where the necessary coupling to auxiliary qubits unavoidably feeds back to the data qubits. Any coupling can then potentially induce errors if the auxiliary qubit or the coupling itself is faulty. A method to extract the error syndrome, which minimizes the interaction between the logical data qubit(s) and the auxiliary qubit(s), was formulated by Steane~\cite{Steane1997}. The key idea is to prepare an auxiliary \emph{logical} qubit using the same code as the data qubit and to couple both logical qubits via a transversal logical CNOT gate,~i.e. in a bit-wise manner. This guarantees that if any single physical operation is faulty, at most one error per encoded logical qubit block is introduced. 
Specifically, first, an auxiliary logical qubit is prepared in a superposition of its basis logical states $|+\rangle_L = \frac{1}{\sqrt{2}}\left(|0\rangle_L + |1\rangle_L\right)$ and a transversal C$^{\mathrm{data}}$NOT$^{\mathrm{aux}}$ is applied, as illustrated in Fig.~\ref{fig:overview_schemes_and_codes}A. In the error-free case, the CNOT will act trivially on both encoded qubits as $\mathrm{CNOT}_L|\psi\rangle_L |+\rangle_L = |\psi\rangle_L |+\rangle_L$. However, if a single bit-flip is present on the $i$th qubit comprising the logical data qubit (denoted as $(X_i^{\mathrm{data}})$), this will be copied onto the $i$th qubit comprising the logical auxiliary qubit (denoted as $(X_i^{\mathrm{aux}})$) as
\begin{align}
    \mathrm{CNOT}_L (X_i^{\mathrm{data}} \otimes& \mathbb{1}) |\psi\rangle_L  \otimes |+\rangle_L  \\
    = X_i^{\mathrm{data}} |\psi\rangle_L &\otimes X_{i}^{\mathrm{aux}} |+\rangle_L. \nonumber
\end{align}
The transversal CNOT gate is FT by construction since it introduces at most one error on each encoded block. The entire circuit is therefore FT if one additionally verifies that only a correctable number of errors is present on the auxiliary logical qubit. The syndrome can then be reconstructed from the outcomes of the projective measurement of the auxiliary logical qubit in a single shot. One can identify the appropriate recovery operation based on a decoder such as the lookup table for the seven-qubit color shown as Tab.~\ref{tab:steane_code_lookuptable} in Appendix~\ref{app:seven_qubit_CC}. Just as an auxiliary logical qubit in $|+\rangle_L$ detects propagated $X$-errors, it detects propagated $Z$-errors when prepared in $|0\rangle_L$ and acted upon with a transversal C$^{\mathrm{aux}}$NOT$^{\mathrm{data}}$ in a second half-cycle. In this second half-cycle $Z$-errors are copied from the data qubit to the auxiliary qubit such that measuring the auxiliary qubits reveals the entire $X$-syndrome simultaneously, just as the first half-cycle reveals the entire $Z$-syndrome.
This Steane-type QEC is to be seen in contrast to measuring each stabilizer individually, where due to fault tolerance requirements one has to resort to either verified Greenberger–Horne–Zeilinger (GHZ) states or flag schemes for the auxiliary qubits~\cite{Shor1996, Chamberland2018flag}. 
Moreover, the syndrome measurement has to be repeated for specific measurement outcomes to avoid single faults leading to high-weight errors, which requires the conditional execution of circuits.
In our experiment and simulation, in order to benchmark against the Steane-type QEC, we implement the flagged syndrome extraction protocol of~\cite{reichardt2020fault}, which was previously realized experimentally in~\cite{ryan2021realization}. The circuits that are used in the implementation are shown in Fig.~\ref{fig:parallel_syndrome_extraction}.

\begin{figure*}[ht]
	\centering
	\includegraphics[width=0.67\linewidth]{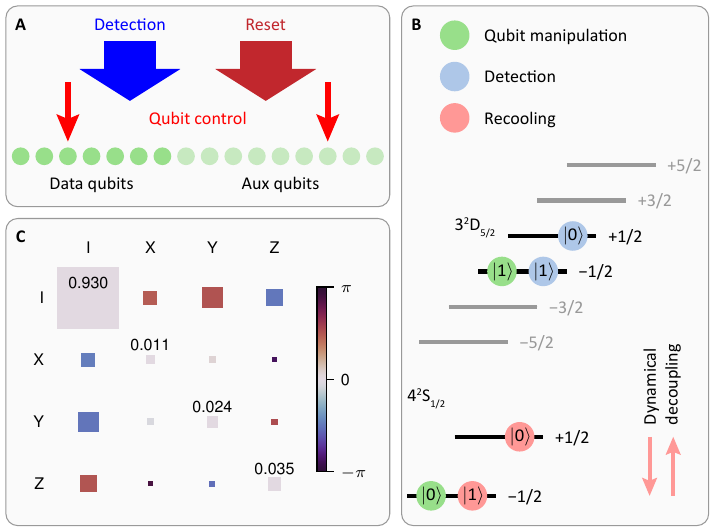}
	\caption{\justifying \textbf{Experimental methods for mid-circuit measurements.} \textbf{A} The qubit register is split into data qubit and auxiliary qubit segments. Tightly focused laser beams addressing up to two individual ions simultaneously are available to manipulate the optical qubit, while qubit state detection and reset lasers illuminate the entire register. (\textbf{B}) Different qubit encodings are used prior to and during different steps of the mid-circuit measurement procedure illustrated by symbols of different coloring. For qubit manipulation the qubits are encoded in two Zeeman-sublevels of two different electronic states (green). The data qubit encoding is transferred to two Zeeman-sublevels of the same electronic state to either decouple from laser light used to project to the computational basis during mid-circuit measurements (blue), or to be able to perform recooling on auxiliary qubits after a mid-circuit measurement (red). While the qubit is encoded in the ground-state manifold (red), we extend the coherence time via dynamical decoupling by applying radio frequency fields. (\textbf{C}) Chi matrix representation~\cite{nielsen2010quantum} of the process acting on data qubits during mid-circuit measurements averaged over all data qubits. The area and the color coding of the squares corresponds to the absolute value and the phase of an element of the chi matrix, respectively. The values on the diagonal of the averaged chi matrix are used to inform the error model (see Appendix~\ref{app:noise_model}).}
	\label{fig:experiment}
\end{figure*}

In this article, we report the implementation of Steane syndrome extraction in a trapped-ion experiment. Central to the implementation of Steane QEC is the transversal logical CNOT, which in our experiment can be performed between all qubits owing to all-to-all qubit connectivity. We employ this to perform single-shot syndrome extraction on different error-correcting codes. As a first step, we investigate the bit-flip and phase-flip repetition codes with code distances $3$ and $5$ each. 
While the repetition code protects only against either Pauli-$X$ or -$Z$ errors, the syndrome extraction procedure is the same as for leading QEC codes such as surface and color codes. We can therefore experimentally explore the scaling of Steane quantum error correction for codes of increasing distance.
Furthermore we demonstrate Steane syndrome extraction for a complete quantum error correcting code by applying it to the seven-qubit color code. We perform up to five and three full cycles of syndrome extraction for the repetition code and seven-qubit color code, respectively. 

\section{Experimental setup}

All experimental results presented in this manuscript are implemented in a trapped-ion quantum processor. Sixteen $^{40}$\textrm{Ca}$^+$ ions are trapped in a macroscopic linear Paul trap, where the electronic state of the ions is controlled via laser pulses, as illustrated in Fig.~\ref{fig:experiment}A. Each ion encodes one qubit in the electronic states $\ket{0}=\ket{4^2\textrm{S}_{1/2}, m_J = -1/2}$ and $\ket{1}=\ket{3^2\textrm{D}_{5/2}, m_J = -1/2}$ (see Fig.~\ref{fig:experiment}B) connected via an optical quadrupole transition at a wavelength of \SI{729}{\nano\meter}. Coulomb interaction between the ions gives rise to collective motional modes of the ions, which are used to mediate entangling operations between any desired pair of qubits. The available universal gate set and its error characteristics are described in more detail in Appendix~\ref{app:experimental_methods}.

For the repeated application of QEC blocks to encoded qubits it is necessary to have the ability to extract the error syndrome by performing measurements on a subset of qubits. Measurements on these auxiliary qubits are designed in a way that minimizes the perturbation of the logical information stored in the data qubits. Furthermore it is beneficial to have the capability of reusing measured qubits by reinitializing them to a defined state in the computational subspace, especially with the limited quantum register sizes of noisy intermediate-scale quantum devices. Viable approaches to implement these procedures in trapped-ion quantum processors are introducing a second atomic species~\cite{negnevitsky2018repeated, pino2021demonstration, erickson2022high}, or moving ions to a distinct region of the trap~\cite{hilder2022fault, zhu2023interactive} for mid-circuit measurements allowing state readout of auxiliary qubits while keeping data qubits unperturbed. In this work we make use of multiple Zeeman-sublevels in the states $\ket{4^2\textrm{S}_{1/2}}$ and $\ket{3^2\textrm{D}_{5/2}}$ (see Fig.~\ref{fig:experiment}) for the implementation of mid-circuit measurements and subsequent reinitialization~\cite{riebe2008deterministic,monz2016realization,manovitz2022trapped}.

\begin{figure*}[t]
	\centering
	\includegraphics[width=0.67\linewidth]{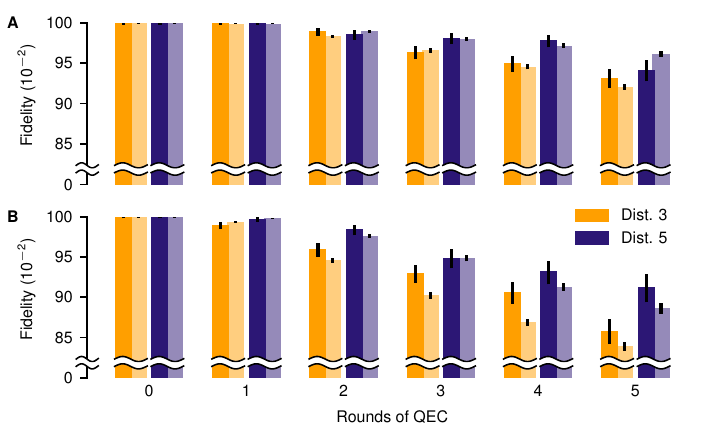}
	\caption{\justifying \textbf{Logical fidelities for bit- and phase-flip repetition code for distances $3$ and $5$.} Fidelities for up to $5$ rounds of Steane-type error correction for the distance-$3$ and distance-$5$ (\textbf{A}) bit-flip and (\textbf{B}) phase-flip code. Round $0$ corresponds to the encoding of the logical state with no extra round of QEC. The experimental and simulation results are depicted with darker and lighter shades, respectively. The error bars shown in this and all following data figures are calculated as described in Appendix~\ref{app:uncertainties}.}
	\label{fig:repetition_code}
\end{figure*}

The first step of this procedure is the detection of the auxiliary qubits by electron shelving~\cite{schindler2013quantum}. All data qubits are encoded in the states shown as blue symbols in Fig.~\ref{fig:experiment}B to retain the phase relation of data qubit superposition states and to prevent scattering out of the computational subspace. Scattering photons from auxiliary qubits projected to $\ket{0}$ heat up the ion string, therefore a Doppler cooling pulse, acting on the same atomic transition also used for auxiliary qubit state detection, is applied. However, further cooling close to the motional ground state using resolved sideband cooling~\cite{schindler2013quantum} is necessary for the implementation of high-fidelity gates after mid-circuit measurements. The sideband cooling procedure involves illumination with laser light that would lead to incoherent relaxation of both states marked with blue symbols to the respective ground states marked as red symbols. Therefore, the data qubit encoding is coherently transferred to the two Zeeman-sublevels of the ground state portrayed as red symbols in Fig.~\ref{fig:experiment}B. Subsequent to sideband cooling a final optical pumping step is used to reinitialize all auxiliary qubits that are supposed to be reused. Finally we restore the encoding of the data qubits to the states shown as green symbols, where further gate operations on the optical qubit can be implemented. A more detailed description of the procedure can be found in Appendix~\ref{app:experimental_methods}.

The duration of the mid-circuit detection procedure is dominated by the sideband cooling step with a duration on the order of the coherence time. Therefore, a dynamical decoupling sequence~\cite{carr1954effects, meiboom1958modified} is performed on the data qubits during the recooling procedure. This decoupling is implemented with a resonant radio frequency antenna driving the transition between the two ground states on the entire register simultaneously, where the data qubits are encoded during sideband cooling (red symbols). A refocusing pulse is applied approximately every millisecond in between cooling pulses for different motional modes. Figure~\ref{fig:experiment}C shows the process matrix~\cite{nielsen2010quantum} of the evolution of data qubits during a full mid-circuit measurement procedure including dynamical decoupling averaged over all data qubits.

\section{Steane QEC for the repetition code and the seven-qubit color code}

\begin{figure*}[t]
	\centering
	\includegraphics[width=0.67\linewidth]{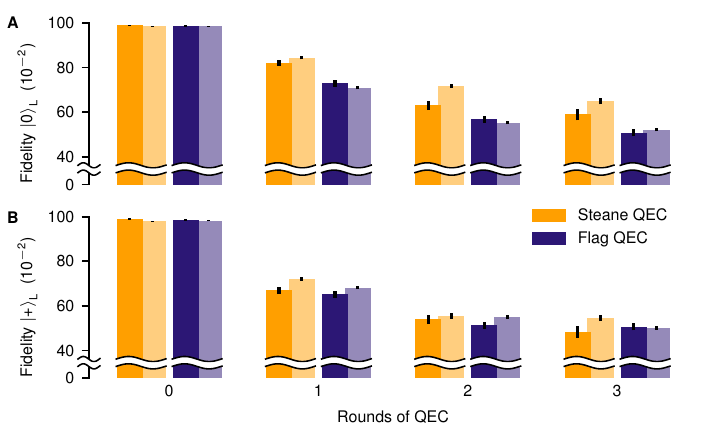}
	\caption{\justifying \textbf{Logical fidelities for syndrome extraction on the seven-qubit color code.} Logical fidelities obtained from Steane-type and flag-based QEC for the logical input state (\textbf{A}) $|0\rangle_L$ and (\textbf{B}) $|+\rangle_L$. Round $0$ corresponds to the encoding of the logical state with no extra round of QEC. The experimental and simulation results are depicted with darker and lighter shades, respectively.}
	\label{fig:comparison_shor_vs_steane}
\end{figure*}

In this section we show the application of Steane QEC to 1D~repetition codes and the 2D seven-qubit color code. First, we study the scaling of Steane QEC with code distance, by presenting results for distances $3$ and $5$ for both the bit- and phase-flip repetition code. The structure of the codes and the stabilizer generators are illustrated in Fig.~\ref{fig:overview_schemes_and_codes}B. Despite their simplicity, these codes share key properties, such as syndrome extraction, logical processing, and error suppression with fully-fledged topological QEC codes. Consequently, they routinely form a testbed for the latter~\cite{Kelly2015,google2021exponential}. One round of error correction in repetition codes consists of only one half of the cycle illustrated in Fig.~\ref{fig:overview_schemes_and_codes}A since the codes can only correct either $X$- or $Z$-errors. 
For example for Steane-type QEC on the distance-$3$ bit-flip code, a logical state $|\psi\rangle_L$ is encoded in $|000\rangle$. 
The auxiliary logical qubit is prepared in a three-qubit GHZ state $|+\rangle_L= (|000\rangle + |111\rangle)/\sqrt{2}$. 
If, for example, an $X$-fault occurs on the first qubit, this will propagate onto the auxiliary logical qubit when the auxiliary qubit is coupled to the logical data qubit with a transversal CNOT gate. A final projective measurement of the auxiliary logical qubit in the $Z$-basis projects the state onto either $|011\rangle$ or $|100\rangle$. 
We can extract the stabilizer values $S_Z^{(1)} = Z_1 Z_2$ and $S_Z^{(2)} = Z_2 Z_3$, as given in Appendix~\ref{app:repetition_code}, by checking the parity of this measurement outcome. It is then possible to identify the initial fault on the first qubit based on the syndrome $(S_Z^{(1)}, S_Z^{(2)}) = (-1, 1)$. 
For the distance-$5$ repetition code, the auxiliary logical state has to be verified to ensure that no single fault has caused a weight-$2$ error configuration on the auxiliary GHZ state. We do this by coupling a single flag qubit to the prepared auxiliary qubit, heralding weight-$2$ error configurations, as shown in Fig.~\ref{fig:repetition_code_circuits} in Appendix~\ref{app:repetition_code}. No verification is required for $d = 3$, since any single fault only results in a correctable error configuration on each encoded block. The phase-flip code can be treated completely analogously to the bit-flip code. 
Figure~\ref{fig:repetition_code} shows the probability to recover the target logical state, within the correction capabilities of the respective QEC code, for the bit- and phase-flip code with distances $d=\{3, 5\}$. We refer to this probability as logical fidelity (see Appendix~\ref{app:logical_fidelity}). Corrections suggested by the repeated syndrome extraction are accounted for via a Pauli frame update~\cite{knill2005quantum}. We can identify that increasing the code distance from $3$ to $5$ improves logical fidelities for both the bit- and the phase-flip code. Lower fidelities for the phase-flip code compared to the bit-flip code can be attributed to dephasing on idling qubits. Experimental results are accompanied by Monte-Carlo simulations using an experimentally informed effective noise model. The model accounts for errors on single-qubit gates, two-qubit gates, qubit initialization, and measurements with error rates $p_{1q} = 3.6\cdot 10^{-3}$, $p_{2q} = 2.7\cdot 10^{-2}$, $p_{\mathrm{init}} = 3\cdot 10^{-3}$ and $p_{\mathrm{meas}} = 3\cdot 10^{-3}$, respectively. Furthermore, during mid-circuit measurements the remaining qubits experience noise which we model as asymmetric depolarizing noise with error probabilities $p^{(x)}_{\mathrm{mid-circ}} = 1.1\cdot 10^{-2}$, $p^{(y)}_{\mathrm{mid-circ}}= 2.4\cdot 10^{-2}$, $p^{(z)}_{\mathrm{mid-circ}}= 3.5\cdot 10^{-2}$, acting on all idling data qubits independently. These error probabilities are extracted from the experimental process matrix, quantifying the effect of mid-circuit measurements on idling (data) qubits, shown in Fig.~\ref{fig:experiment}C. A more detailed description of the error model can be found in Appendix~\ref{app:noise_model}. The simulation data obtained with this relatively simple multi-parameter, incoherent noise model shows good agreement with the experimental data, which indicates that it captures the experiment well for the given error rates.

Going a step further, we now apply the above-described Steane-type QEC to the seven-qubit color code~\cite{steane1996multiple, bombin2006topological, nigg2014quantum}, shown in Fig.~\ref{fig:overview_schemes_and_codes}C, and compare its performance to the previously used flag-based QEC scheme~\cite{reichardt2020fault, ryan2021realization, postler2022demonstration}. In the latter, the syndrome information is extracted by measuring stabilizers using additional auxiliary physical qubits, as illustrated in Fig.~\ref{fig:overview_schemes_and_codes}D.
The seven-qubit color code $[[7, 1, 3]]$ is the smallest topological color code and encodes a single logical qubit while allowing the correction of a single arbitrary Pauli error. It has the highly desirable property of admitting a transversal and thus FT implementation of the entire Clifford group. Physical qubits are placed on the vertices of a two-dimensional graph and the encoded logical qubit is defined as the simultaneous $+1$-eigenstate of the six indicated stabilizer generators. A single flag qubit is used to verify the prepared logical state such that unsuccessful preparations can be discarded~\cite{goto2016minimizing, postler2022demonstration} analogous to usage in the distance-$5$ repetition code. The encoding circuit and lookup table are given in Appendix~\ref{app:seven_qubit_CC}. 

Figure~\ref{fig:comparison_shor_vs_steane} shows the logical fidelities we obtained experimentally and numerically from Monte-Carlo simulations. We find an advantageous performance in terms of fidelity for Steane-type QEC compared to the flag-based QEC scheme, where the performance benefit of Steane QEC is more pronounced for the state $|0\rangle_L$. The reason for this asymmetry is that the dominating error source in the experimental setup at hand is asymmetric depolarizing noise on the data qubits during mid-circuit measurements. Additionally dephasing of data qubits is taking place during the implementation of gates on the auxiliary qubits (see Tab.~\ref{tab:error_rates_simulation} in Appendix~\ref{app:noise_model}). This conflates the logical failure rates for the different protocols and partially veils the advantage of Steane QEC, which is most pronounced in the regime of dominating two-qubit error rates. Therefore, we estimate the projected advantage of Steane-type QEC by numerically simulating logical error rates in a setting where the only noise source are two-qubit gate errors. In this regime, we find that the logical error rate of Steane-type QEC is suppressed by as much as a factor of 2, compared to flag-type QEC (see Fig.~\ref{fig:limiting_case_syndrome_extractions} in Appendix~\ref{app:seven_qubit_CC}). Experimental improvements like an extended coherence time on the order of seconds~\cite{harty2014high, ruster2016long, wang2021single} or composite pulses robust against laser amplitude noise and crosstalk~\cite{wimperis1989composite,wimperis1994broadband} could further mitigate perturbations of idling data qubits during mid-circuit measurements and therefore extend the advantage in logical fidelity offered by the Steane-type over flag-type QEC also in the present experimental setup. Additional results on the extraction of only the $Z$ ($X$) syndrome for the logical state $|0\rangle_L$ ($|+\rangle_L$) are presented in Appendix~\ref{app:halfcycles}.

\section{Conclusions and Outlook}
In this work we show practical advantage of Steane over flag-based QEC in a noisy intermediate-scale trapped-ion quantum processor. We have implemented up to five rounds of Steane QEC for bit-flip and phase-flip repetition codes with distances $3$ and $5$, and observe an improvement of the logical fidelity with larger distances. This increase in spite of larger qubit and gate overhead per logical qubit shows that both codes were operated below their respective thresholds. We further demonstrated an advantage of repeated Steane QEC on the seven-qubit color code, where multiple complete rounds of error correction including repeated mid-circuit measurements present substantial experimental challenges. Numerical simulations based on a multi-parameter depolarizing error model, informed by experimentally estimated error rates of basic quantum operations, underpin this finding and capture the features in the experimental data. The improved QEC performance has been achieved without the necessity to make any changes to the hardware, but is rooted in the carefully crafted quantum circuit design underlying the Steane-type QEC approach. The present implementation is currently limited by errors during mid-circuit measurements. Therefore, the benefit of Steane QEC will increase up to the numerically anticipated margin of about a factor of 2, if the error rate of the mid-circuit measurement procedure becomes smaller than the entangling gate errors.

The results presented in this work establish Steane QEC as a new paradigm in experimental QEC by showing reduced error rates of encoded qubits compared to other QEC protocols.
The demonstrated Steane-type QEC approach is especially relevant in the context of the emergence of larger qubit registers with efficient implementations of entangling logical gates on various platforms. The possibility of applying transversal CNOT gates in parallel in this paradigm enables the extraction of $Z$-type and $X$-type error syndromes each within one circuit time step.
We believe Steane QEC will play a pivotal role towards large-scale fault-tolerant quantum computation owing to its increased logical fidelities, and modularity allowing to harness the emerging capabilities of efficiently and fault-tolerantly coupled logical qubits as building blocks.

\section*{Acknowledgments}

Please also note the preprint entitled ‘Comparing Shor and Steane Error Correction Using the Bacon-Shor Code’ by S. Huang, K. R. Brown, and M. Cetina on related work. 

\subsection*{Funding:}
We gratefully acknowledge support by the European Union’s Horizon Europe research and innovation program under Grant Agreement Number 101114305 (“MILLENION-SGA1” EU Project), the US Army Research Office through Grant Number W911NF-21-1-0007, the European Union’s Horizon Europe research and innovation program under Grant Agreement Number 101046968 (BRISQ), the ERC Starting Grant QCosmo under Grant Number 948893, the ERC Starting Grant QNets through Grant Number 804247, the Austrian Science Fund under Project Number F7109 (SFB BeyondC), the Austrian Research Promotion Agency under Contracts Number 896213 (ITAQC), the Office of the Director of National Intelligence (ODNI), Intelligence Advanced Research Projects Activity (IARPA), via the U.S. Army Research Office through Grant Number W911NF-16-1-0070 under the LogiQ program, and by IARPA and the Army Research Office, under the Entangled Logical Qubits program through Cooperative Agreement Number W911NF-23-2-0216. This research is also part of the Munich Quantum Valley (K-8), which is supported by the Bavarian state government with funds from the Hightech Agenda Bayern Plus.
We further receive support from the IQI GmbH, and by the German ministry of science and education (BMBF) via the VDI within the project IQuAn, and by the Deutsche Forschungsgemeinschaft (DFG, German Research Foundation) under Germany’s Excellence Strategy ‘Cluster of Excellence Matter and Light for Quantum Computing (ML4Q) EXC 2004/1’ 390534769.

The views and conclusions contained in this document are those of the authors and should not be interpreted as representing the official policies, either expressed or implied, of IARPA, the Army Research Office, or the U.S. Government. The U.S. Government is authorized to reproduce and distribute reprints for Government purposes notwithstanding any copyright notation herein.

\subsection*{Authors contribution:}
L.P. and I.P. carried out the experiments. L.P., I.P., C.D.M., P.S. and T.M. contributed to the experimental setup. L.P., F.B. and M.R. analyzed the data. F.B., S.H. and M.R. performed the numerical simulations. F.B., S.H. and M.R. performed circuit analysis, characterization and theory modeling. L.P., F.B., I.P., C.D.M., S.H., P.S. and M.R. wrote the manuscript, with contributions from all authors. R.B., P.S., M.R., M.M. and T.M. supervised the project.

\subsection*{Competing interests:}
R.B. and T.M. are connected to Alpine Quantum Technologies GmbH, a commercially oriented quantum computing company.

\subsection*{Data and materials availability:}
The data underlying the findings of this work are available at https://doi.org/10.5281/zenodo.10390470.
All codes used for data analysis are available from the corresponding author upon reasonable request.

\bibliography{bib}
\bibliographystyle{apsrevmod}

\onecolumngrid
\newpage
\appendix
\renewcommand{\thefigure}{A\arabic{figure}}
\setcounter{figure}{0}

\section{Quantum error correcting codes}

\subsection{The seven-qubit color code}\label{app:seven_qubit_CC}
\begin{figure}[ht]
	\centering
	\includegraphics[width=0.33\linewidth]{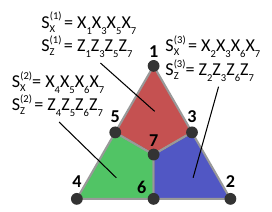}
	\caption{\justifying \textbf{Stabilizer generators of the seven qubit color code.} The seven-qubit color code encodes one logical qubit into seven physical qubits. A code state is a $+1$ eigenstate of all six weight-$4$ stabilizer generators \{$S^{(i)}_X, S^{(i)}_Z$\} defined on the colored plaquettes. Pauli-$X$ (-$Z$) on all qubits corresponds to the logical operator $X_L$ ($Z_L$), which is up to multiplication with stabilizer generators equivalent to weight-$3$ operators.}
	\label{fig:colorcode_standalone}
\end{figure}

\begin{figure}
	\centering
	\includegraphics[width=0.5\linewidth]{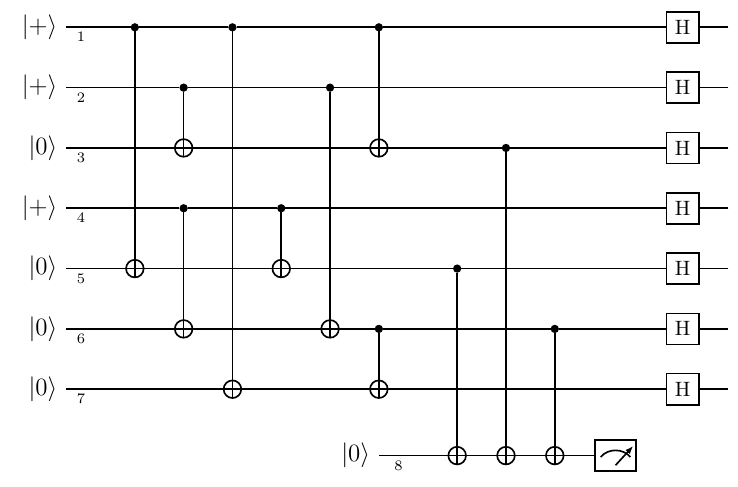}
	\caption{\justifying \textbf{Circuit for the encoding of a logical state in the seven-qubit color code~\cite{goto2016minimizing, postler2022demonstration}}. The first eight CNOT gates initialize $|0\rangle_L$ on the seven-qubit color code. This is followed by a verification in order to detect single faults that would otherwise propagate onto multiple data qubits and cause a logical failure. Finally, a transversal application of $H_L$ may be applied to prepare $|+\rangle_L$. }
	\label{fig:encoding_steane_code}
\end{figure}

The seven-qubit color code $[[7, 1, 3]]$ is constructed by placing physical qubits on the vertices of a graph~\cite{steane1996multiple}. The encoded logical qubit is defined as the simultaneous $+1$-eigenstate of the six stabilizer generators 
\begin{align}
    S^{(1)}_X = X_1 X_3 X_5 X_7,  \quad S^{(1)}_Z = Z_1 Z_3 Z_5 Z_7 \nonumber\\
    S^{(2)}_X = X_4 X_5 X_6 X_7,  \quad S^{(2)}_Z = Z_4 Z_5 Z_6 Z_7 \\
    S^{(3)}_X = X_2 X_3 X_6 X_7,  \quad S^{(3)}_Z = Z_2 Z_3 Z_6 Z_7,  \nonumber
\end{align}
as illustrated in Fig.~\ref{fig:overview_schemes_and_codes}A and Fig.~\ref{fig:colorcode_standalone}. The logical operators are given by $Z_L = Z^{\otimes 7}$ and $X_L=X^{\otimes 7}$, which can be expressed as weight-$3$ operators by multiplication with stabilizers. For instance, multiplying $S^{(1)}_X$ with $X_L$ gives the weight-$3$ logical operator $X_2 X_4 X_6$. The circuit shown in Fig.~\ref{fig:encoding_steane_code} is used to encode a logical state in the seven-qubit color code~\cite{goto2016minimizing}. 

In the error-free case, a measurement of the set of stabilizer generators will yield the outcome $+1$ for each one, since a valid code state is a $+1$-eigenstate of these operators. 
If a single Pauli fault occurs, this will anticommute with a set of stabilizers. 
The measurement outcomes in this case will yield the outcome $-1$ for a set of stabilizers. This syndrome measurement outcome is unique to the initial single Pauli fault when excluding error configurations of weight greater than one. Therefore, one can correct for this error and recover the code state. However, higher-weight error configurations break this uniqueness and can lead to logical errors when applying the recovery operation. For instance, the weight-$1$ error configuration $X_5$ and the weight-$2$ configuration $X_1 X_4$ lead to the same $Z$ syndrome $\{S^{(1)}_Z=-1, S^{(2)}_Z=-1, S^{(3)}_Z=1\}$. Applying the recovery operation $X_5$ would lead to a logical error for the weight-$2$ error case, as the weight-$3$ configuration $X_1 X_4 X_5 = X_L S^{(3)}_X$ is up to multiplication with a stabilizer generator equivalent to $X_L$.
Table~\ref{tab:steane_code_lookuptable} summarizes the possible $Z$-syndrome measurement outcomes and the corresponding recovery operation, which corrects any single Pauli $X$-error. Since the seven-qubit color code is self-dual, i.e. symmetric under exchange of $X-$ and $Z-$stabilizers, the lookup table for Pauli $Z$-corrections based on the measured $X$-syndrome is the same. 

\begin{figure}
	\centering
	\includegraphics[width=1.\linewidth]{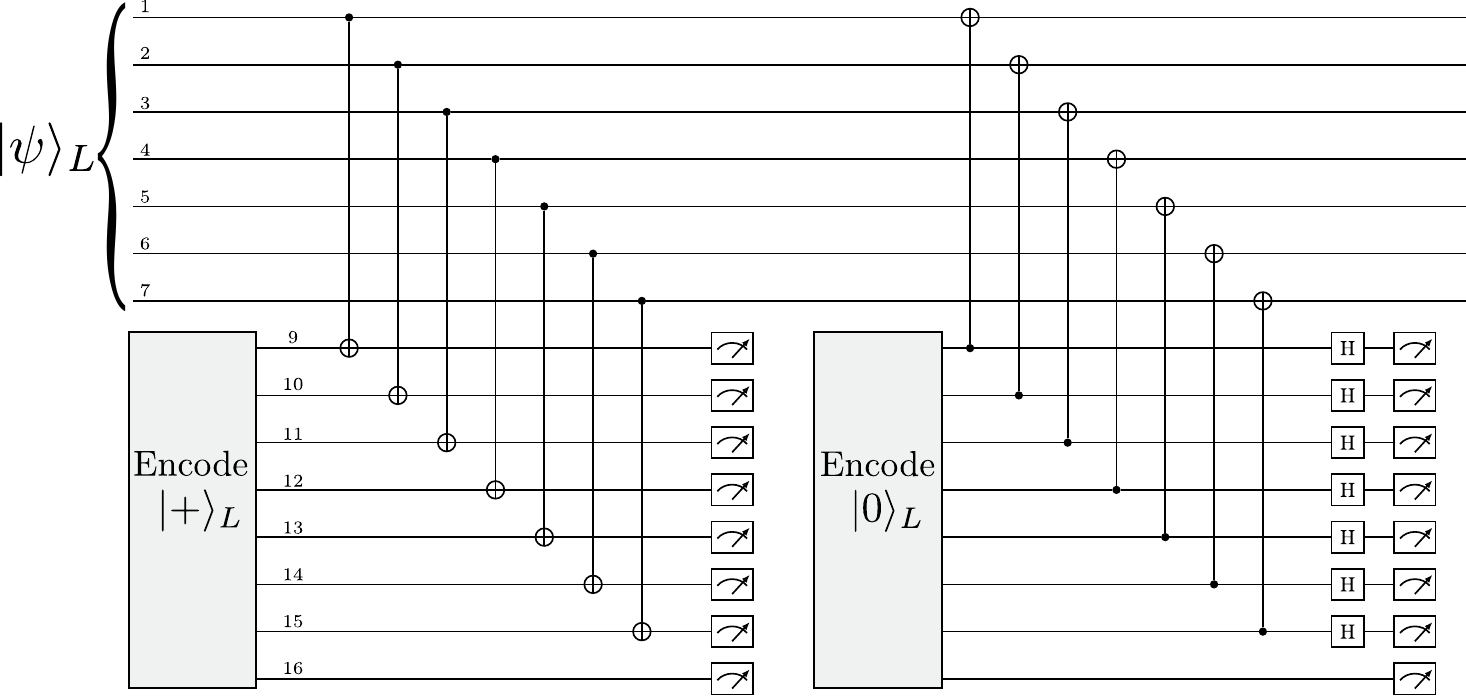}
	\caption{\justifying \textbf{Circuits for Steane-type syndrome extraction in the seven-qubit color code~\cite{Steane1997, aliferis2006quantum, chamberland2018new}}. A logical auxiliary qubit is initialized using the circuit shown in Fig.~\ref{fig:encoding_steane_code}, coupled to the data qubits and measured.}
	\label{fig:steane_EC_circuit}
\end{figure}

\begin{figure}
	\centering
	\includegraphics[width=1.\linewidth]{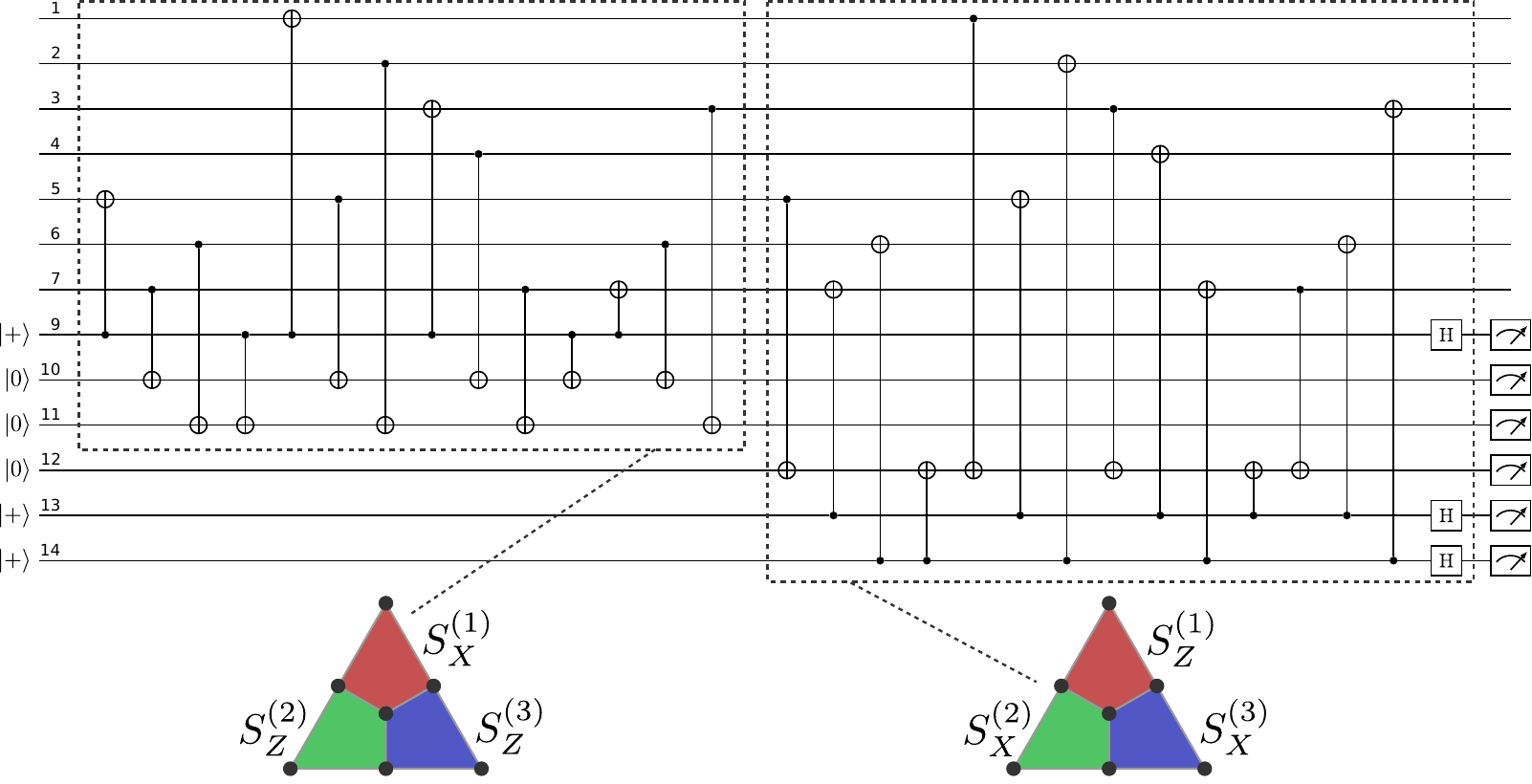}
	\caption{\justifying \textbf{Circuit for flagged syndrome readout for the syndrome extraction on the seven-qubit color code~\cite{ryan2021realization, postler2022demonstration, reichardt2020fault}}. The first part of the circuit measures $S_X^{(1)}, S_Z^{(2)}, S_Z^{(3)}$ and the second part extracts $S_Z^{(1)}, S_X^{(2)}, S_X^{(3)}$. }
	\label{fig:parallel_syndrome_extraction}
\end{figure}

The syndrome can be extracted using Steane-type QEC with the circuit shown in Fig.~\ref{fig:steane_EC_circuit}, where transversal CNOT gates copy errors onto an auxiliary logical qubit which is then measured projectively. Figure~\ref{fig:parallel_syndrome_extraction} shows the circuit we use for flagged syndrome readout on the seven-qubit color code~\cite{ryan2021realization, postler2022demonstration, reichardt2020fault}. If no error is detected in a first round of flagged stabilizer measurements, we assume that no error has occurred and proceed. If a non-trivial syndrome is measured, the complete syndrome is measured again with unflagged circuits to distinguish the dangerous propagated flag errors from non-flag errors.
If the two syndromes agree, we take this as a final syndrome for error correction with lookup table Tab.~\ref{tab:steane_code_lookuptable}. If they do not agree and the unflagged syndrome coincides with a flag-error syndrome in Tab.~\ref{tab:steane_code_flaglookuptable}, we apply the corresponding flag-error correction. In case the unflagged syndrome is not in the flag-lookup table, the single-qubit recovery from Tab.~\ref{tab:steane_code_lookuptable} is used. Note that while one could immediately apply the correction to the data register, it is admissible to just keep track of this in software (known as Pauli frame tracking) as long as no logical non-Clifford gate is applied. The experimental setup currently does not allow for real-time changes of the gate sequence based on outcomes of mid-circuit measurements. For the realization of flagged QEC we experimentally implement both possible circuits, with and without a second unflagged measurement of the stabilizer generators, and all combinations thereof for multiple QEC cycles. In post-selection we discard all implementations where the flagged syndrome was trivial, but an unflagged readout was following, and vice-versa.

\begin{table}[ht]
    \centering
    \renewcommand*{\arraystretch}{1.1}
    \begin{tabular}{c|c}
         $S_Z^{(1)}, S_Z^{(2)}, S_Z^{(3)}$ & Recovery \\
         \hline
         $+++$ & $I$\\
         \hline
         $-++$ & $X_1$\\
         \hline
         $++-$ & $X_2$\\
         \hline
         $-+-$ & $X_3$\\
         \hline
         $+-+$ & $X_4$\\
         \hline
         $--+$ & $X_5$\\
         \hline
         $+--$ & $X_6$\\
         \hline
         $---$ & $X_7$
    \end{tabular}
    \caption{\justifying \textbf{Lookup table for the seven-qubit color code.} For each syndrome measurement with a given set of positive $+$ and negative $-$ outcomes, a single qubit recovery operation is applied. Since $X$- and $Z$-stabilizers are defined symmetrically on the same support, the $Z$-type recoveries based on the $X$-syndrome are applied analogously. Any $Y$-error can be considered as a combined $X$- and $Z$-error and be corrected independently.}
    \label{tab:steane_code_lookuptable}
\end{table}

\begin{table}[ht]
    \centering
    \renewcommand*{\arraystretch}{1.1}
    \begin{tabular}{c|c}
         $S_Z^{(1)}, S_Z^{(2)}, S_Z^{(3)}$ & Recovery \\
         \hline
         $+-+$ & $X_3X_7$\\
         \hline
         $++-$ & $X_4X_6$
    \end{tabular}
    \caption{\justifying \textbf{Flag-lookup table for the seven-qubit color code.} If the outcomes of flagged and unflagged syndrome readouts do not agree and the measured unflagged syndrome is the table above, the corresponding recovery operation is applied. If the unflagged syndrome is not in the above table, the respective single-qubit recovery operation from Tab.~\ref{tab:steane_code_lookuptable} is applied. Since $X$- and $Z$-stabilizers are defined symmetrically on the same support, the $Z$-type recoveries based on the $X$-syndrome are applied analogously.}
    \label{tab:steane_code_flaglookuptable}
\end{table}

\begin{figure}
	\centering
	\includegraphics[width=0.67\linewidth]{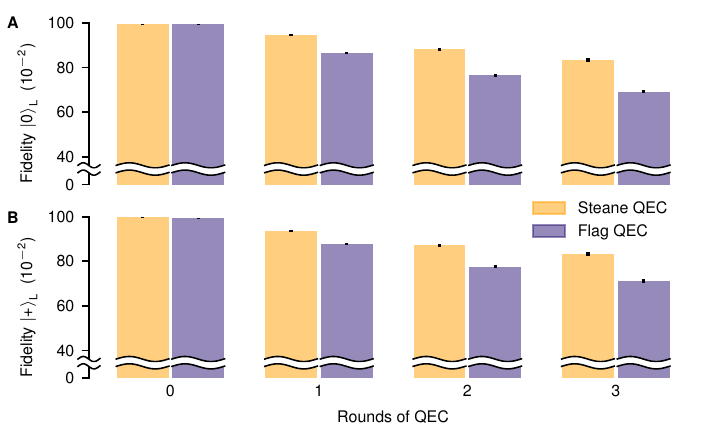}
	\caption{\justifying \textbf{Fidelities from numerical simulations for Steane-type and flagged syndrome extraction in the limiting case of only accounting for two-qubit gate errors.} All error probabilities, apart from the two-qubit gate error probability $p_{2q} = 0.025$, are set to $p_{1q} = p_{\mathrm{init}} = p_{\mathrm{meas}} = p_{\mathrm{mid-circ}} = 0$. Shown are the logical fidelities for the seven-qubit color code after preparing the logical states (\textbf{A}) $|0\rangle_L$ and (\textbf{B}) $|+\rangle_L$. Steane-type QEC reaches higher fidelities than the flag-based approach and this difference increases with the number of subsequent rounds of QEC. Round $0$ corresponds to the encoding of the logical state with no extra round of QEC.}
	\label{fig:limiting_case_syndrome_extractions}
\end{figure}

We numerically calculate the fidelities for the limiting case where all error rates except $p_{2q}$ are set to $0$, as shown in Fig.~\ref{fig:limiting_case_syndrome_extractions}, in order to estimate the potential advantage of Steane-type QEC over the flag-based approach. Since there is no additional dephasing included, the systematic difference in fidelity between the two logical states $|0\rangle_L$ and $|+\rangle_L$ vanishes. 
The fidelities for the Steane-type approach are higher than for the flag-based protocol and this difference increases with the number of QEC rounds. After two rounds of QEC, the fidelity for the Steane-type approach is already more than $0.1$ higher than for the flag-based protocol. 
This promises an advantage of Steane-type QEC in the regime of dominating two-qubit error rates. 

\subsection{The 1D repetition code}\label{app:repetition_code}
\begin{figure}
	\centering
	\includegraphics[width=1.\linewidth]{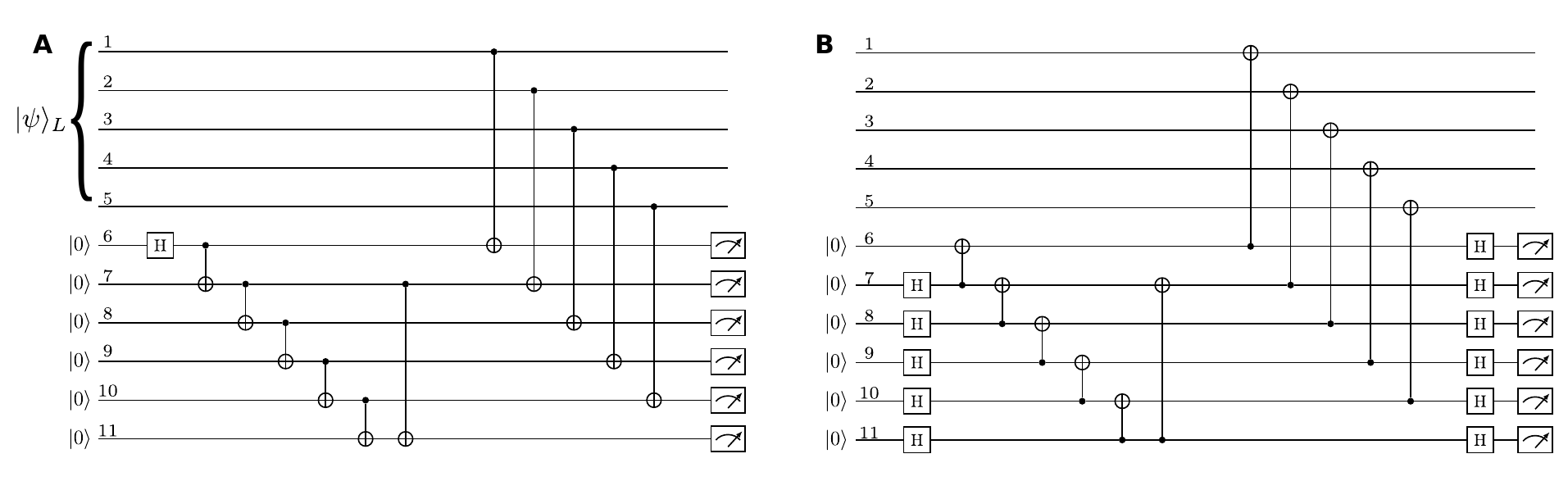}
	\caption{\justifying \textbf{Circuits for Steane-type syndrome extraction in the repetition code.} Syndrome extraction for the \textbf{(A)} bit-flip code or \textbf{(B)} phase-flip code is performed by initializing a logical auxiliary qubit is in a suitable GHZ-state on qubits $6$ to $10$ and verifying it with an additional flag qubit $11$. Sequentially, a transversal CNOT gate is applied and the auxiliary qubit is measured in the corresponding basis. The flag verification (qubit 11) can be left out in the circuits for the respective distance-$3$ codes without breaking fault tolerance. }
\label{fig:repetition_code_circuits}
\end{figure}

For the $n$-qubit bit-flip code~\cite{nielsen2010quantum}, the logical $|0\rangle_L$ is encoded in $n$ copies of $|0\rangle$ as $\ket{0}_L = \ket{0}^{\otimes n}$. 
The stabilizer generators are given by pairs of neighboring Pauli-$Z$ operators \{$Z_1 Z_2, Z_2Z_3, ...Z_{n-1}Z_{n}$\} and the logical operators are $X_L = X^{\otimes n}$ and $Z_L = Z_1$. 
Analogously, the phase-flip code takes repetitions of $|+\rangle$ to encode information redundantly on multiple qubits. In this case, one can define the $n$-qubit state $|0\rangle_L = |+\rangle^{\otimes n}$ and the stabilizers correspond to pairs of Pauli-$X$-operators \{$X_1 X_2, X_2X_3, ...X_{n-1}X_n$\} and the logical operators are given by $X_L = Z^{\otimes n}$ and $Z_L = X_1$. Steane-type QEC is performed by initializing a second logical qubit in the corresponding logical $|+\rangle_L$ and applying a transversal CNOT gate, as shown exemplarily in Fig.~\ref{fig:repetition_code_circuits} for the distance-$5$ repetition code. 

\section{Effective noise model and simulation methods}\label{app:noise_model}
In order to estimate the logical fidelities of the discussed error correction protocols, we perform Monte-Carlo simulations using STIM~\cite{Gidney2021} and PECOS~\cite{ryan2018pecos}. 
Every component in a circuit is modeled as an ideal operation $U_{\mathrm{ideal}}$ followed by an error $E$ drawn from an error set with a given probability $p$. We consider depolarizing noise channels on single- and two-qubit gates 
\begin{align}
    \mathcal{E}_1(\rho) &= (1 - \sum_{j=1}^3 p_j)\rho + \sum_{j= 1}^3 p_j E^{j}_1 \rho E^{j}_1 \\
    \mathcal{E}_2(\rho) &= (1 - p_{2q})\rho + \frac{p_{2q}}{15} \sum_{j= 1}^{15}   E_2^{j} \, \rho\, E_2^{j}.  \nonumber\label{eq:depol_single_qubit}
\end{align}
with the error sets 
\begin{align}
	E_1 &\in \{ \sigma_k, \forall k \in \{1, 2, 3 \} \} \\
	E_2 &\in \{\sigma_k \otimes \sigma_l, \forall k, l \in \{0, 1, 2, 3 \}  \} \backslash  \{\sigma_0 \otimes \sigma_0 \}, \nonumber
\end{align}
where $\sigma_k$ are the single-qubit Pauli operators $\sigma_k = \{I, X, Y, Z \}$ with $k=0, 1, 2, 3$. 
For the single qubit case we give the more general channel to capture the asymmetric case found for the errors induced during mid-circuit measurements. The general formula reduces to the symmetric depolarizing case by choosing all three Pauli errors with equal probability $p_j=p_{1q}/3$, as we do for the single-qubit gate error. The parameters $p_{1q}$ and $p_{2q}$ specify the probability that any one of the errors of the corresponding error set occurs on the qubits which are acted upon. Qubits are initialized and measured in the computational basis. Faults on these operations are modeled by applying $X$-errors after state preparations and before measurements with a probability $p_{\mathrm{init}}$ and $p_{\mathrm{meas}}$, respectively. Furthermore, we include noise on idling qubits, which are not acted upon with a gate at a given step of the protocol. Since the dominating noise on idling qubits is dephasing due to magnetic field fluctuations, which limits the decoherence time $T_2 \approx 50\,$ms, we model the noise channel for idling qubits by Pauli-$Z$ faults as
\begin{align}
    \mathcal{E}_{\mathrm{idle}}(\rho) &= (1 - p_{\mathrm{idle}})\rho + p_{\mathrm{idle}} Z\rho Z. 
\end{align}
The probability $p_{\mathrm{idle}}$ of a $Z$-fault on each idling qubit depends on the execution time $t$ of the performed gate and is given by
\begin{align}
    p_{\mathrm{idle}} = \frac{1}{2} \left[1 - \mathrm{exp}\left(-\frac{t}{T_2} \right) \right]. 
\end{align}

Mid-circuit detections are performed in order to perform multiple rounds of error correction, where auxiliary qubits are measured while keeping the data qubits intact. The idling data qubits experience noise during this mid-circuit detection, which we model as an asymmetric depolarizing channel on all data qubits. We estimate the individual Pauli $p^{(x)}_{\mathrm{mid-circ}}, p^{(y)}_{\mathrm{mid-circ}}, p^{(z)}_{\mathrm{mid-circ}}$ error rates in this channel based on single-qubit process tomography. All error rates and gate times are summarized in Tab.~\ref{tab:error_rates_simulation}. 

\begin{table}
    \centering
    \renewcommand*{\arraystretch}{1.1}
    \begin{tabular}{c|c|c}
         Operation & Error rate & Duration\\
         \hline
         Two-qubit gate & $p_{2q} = 0.027$ & \SI{322.5}{\micro s}\\
         \hline
         Single-qubit gate & $p_{1q} = 0.0036$ & \SI{25}{\micro s}\\
         \hline
         Measurement & $p_{\mathrm{meas}} = 0.003$& \\
         \hline
         Preparation & $p_{\mathrm{init}} = 0.003$& \\
         \hline
         & $p^{(x)}_{\mathrm{mid-circ}}= 0.011$ & \\ 
         Mid-circuit detection  & $p^{(y)}_{\mathrm{mid-circ}}= 0.024$ & \\ 
         & $p^{(z)}_{\mathrm{mid-circ}}= 0.035$ & \\
    \end{tabular}
    \caption{\justifying \textbf{Error rates and duration of operations on a trapped-ion quantum processor. } These values correspond to the trapped-ion setup that was used in the experiments and are used in the following simulations. Gate durations are increased by \SI{10}{\micro\second} compared to the values given in Appendix~\ref{app:experimental_methods} to account for settling times of the addressing optics, when the ion being addressed is changed.}
    \label{tab:error_rates_simulation}
\end{table}

\section{Experimental methods}\label{app:experimental_methods}

\begin{figure}
    \centering
    \includegraphics[width=0.67\linewidth]{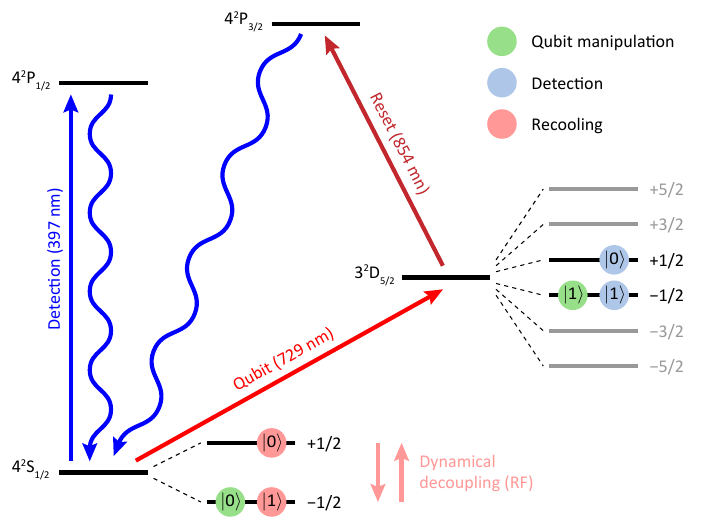}
    \caption{\justifying \textbf{Energy scheme of $^{40}$\textrm{Ca}$^+$.} Laser at wavelengths of \SI{729}{\nano\meter}, \SI{854}{\nano\meter} and \SI{397}{\nano\meter} are used for qubit manipulation, qubit reset, and Doppler cooling, respectively. Different encodings are used for the data qubits in different stages of a circuit. Gates are applied to the optical qubit encoding (green symbols). For mid-circuit measurements the data qubits are encoded in different Zeeman-sublevels of the states $\ket{0}=\ket{4^2\textrm{S}_{1/2}}$ and $\ket{0}=\ket{3^2\textrm{D}_{5/2}}$ to avoid interaction with laser light for qubit state detection, reset, and recooling illuminating the whole register. Furthermore a radio frequency drive is available for dynamical decoupling when the qubit is encoded in the Zeeman-manifold of the electronic ground state during the sideband recooling stage of mid-circuit measurements.}
    \label{fig:setup_detailed}
\end{figure}

All experiments presented in this manuscript are conducted on a trapped-ion quantum processor~\cite{Pogorelov2021}. In a non-segmented macroscopic Paul trap a string of sixteen $^{40}$\textrm{Ca}$^+$ ions is trapped with inter-ion spacing ranging from \SI{3.8}{\micro\meter} to \SI{6.0}{\micro\meter}, set by trap parameters. The center-of-mass modes of the ion string in the pseudo-harmonic potential of the trap have oscillation frequencies of $\omega_z = 2\pi \times \SI{369}{\kilo\hertz}$ and $\omega_x = 2\pi \times \SI{3086}{\kilo\hertz}$, $\omega_y = 2\pi \times \SI{3165}{\kilo\hertz}$ for the direction along the ion string (referred to as axial) and the two perpendicular directions (referred to as radial), respectively. The first step of every experimental cycle is a Doppler cooling pulse acting on the $\ket{4^2\textrm{S}_{1/2}} \leftrightarrow \ket{4^2\textrm{P}_{1/2}}$ transition at a wavelength of \SI{397}{\nano\meter} (see Fig.~\ref{fig:setup_detailed}) with a duration of \SI{500}{\micro\second}. Simultaneously the ion chain is illuminated with light at \SI{866}{\nano\meter} to avoid pumping to the dark state $\ket{3^2\textrm{D}_{3/2}}$ by driving any population trapped there back to the state $\ket{4^2\textrm{S}_{1/2}}$. Both laser beams act on all ions simultaneously and have spatial overlap with all motional modes. The two lowest-frequency axial modes and all $32$ radial modes are further cooled close to the ground state by resolved sideband cooling~\cite{schindler2013quantum}. The ions are illuminated by a laser beam red-detuned by the respective motional frequency from the $\ket{4^2\textrm{S}_{1/2}, m_J = -1/2} \leftrightarrow \ket{3^2\textrm{D}_{5/2}, m_J = -5/2}$ transition at \SI{729}{\nano\meter}. For the axial modes the \SI{729}{\nano\meter} laser propagates along the ion string, whereas the radial modes are cooled by a steerable, addressed \SI{729}{\nano\meter} beam illuminating only one ion at a time from a direction perpendicular to the ion string. To cool the radial modes the ion having the strongest coupling to the respective motional mode is illuminated. All modes within approximately \SI{50}{\kilo\hertz} from the laser frequency are cooled, which allows us to cool 34 modes with only 15 frequency settings. To accelerate the cooling process ions excited to the $\ket{3^2\textrm{D}_{5/2}}$ state are pumped to the state $\ket{4^2\textrm{P}_{3/2}}$ using laser light at \SI{854}{\nano\meter} from where they rapidly decay to one of the ground states. This sideband cooling cycle is repeated up to five times depending on the mode, where one cycle takes \SI{500}{\micro\second}. The finalizing step of the state preparation procedure is to prepare all ions in $\ket{4^2\textrm{S}_{1/2}, m_J = -1/2}$ by exciting the transition $\ket{4^2\textrm{S}_{1/2}, m_J = +1/2} \leftrightarrow \ket{3^2\textrm{D}_{5/2}, m_J = -3/2}$ with axial \SI{729}{\nano\meter} light while speeding up the decay to the ground state using \SI{854}{\nano\meter} light, as for sideband cooling.

The coherent manipulation of the individual qubits in the register subsequent to state preparation is exclusively done via tightly focused laser pulses with a propagation direction perpendicular to the ion string addressing the qubit transition $\ket{4^2\textrm{S}_{1/2}, m_J = -1/2} \leftrightarrow \ket{3^2\textrm{D}_{5/2}, m_J = -1/2}$. The experimental setup allows us to illuminate up to two ions simultaneously. Addressing a single ion $i$ with light resonant to the qubit transition allows us to implement operations of the form $R_\phi^{(i)}(\theta) = \exp{(-\mathrm{i}\frac{\theta}{2}(\sigma_x^{(i)}\cos{\phi} + \sigma_y^{(i)}\sin{\phi}))}$, where $\sigma_\bullet$ are single-qubit Pauli matrices. The rotation axis $\phi$ can be controlled via the light phase, the rotation angle $\theta$ via light intensity and pulse duration. The duration of a pulse with $\theta = \frac{\pi}{2}$ is \SI{15}{\micro\second}. Together with virtual Z rotations~\cite{mckay2017efficient} this operation allows us to implement arbitrary single-qubit unitary operations. The gate set is completed by adding the entangling M{\o}lmer-S{\o}rensen interaction, where any two ions may be illuminated with bichromatic light slightly detuned from the motional sidebands corresponding to the radial mode at frequency $\omega_y$~\cite{sorensen2000entanglement}. Adjusting the gate time to $t_{\mathrm{gate}} = \frac{1}{\delta} =$ \SI{312.5}{\micro\second}, where $\delta$ is the detuning from the motional sidebands, implements the gate $\mathrm{MS}^{(ij)} = \exp({-\mathrm{i}\frac{\pi}{4}\sigma_x^{(i)}\sigma_x^{(j)}})$, which is equivalent to the CNOT gate up to single-qubit rotations~\cite{maslov2017basic}.
Qubit state readout after application of a gate sequence to the qubit register is implemented by simultaneously illuminating the whole register with laser light at wavelengths of \SI{397}{\nano\meter} and \SI{866}{\nano\meter}. While qubits projected to the computational state $\ket{0}=\ket{4^2\textrm{S}_{1/2}, m_J = -1/2}$ repeatedly emit photons at a wavelength of \SI{397}{\nano\meter} after being excited to $\ket{4^2\textrm{P}_{1/2}}$ and returning to $\ket{4^2\textrm{S}_{1/2}}$, qubits projected to $\ket{1}=\ket{4^2\textrm{D}_{5/2}, m_J = -1/2}$ are not affected by those light fields and do not emit photons~\cite{schindler2013quantum}. Imaging the ion string on an electron multiplying charge-coupled device (EMCCD) camera allows for the spatially resolved detection of light emission from the ion string, and therefore the computational basis bit string the qubit register was projected into can be reconstructed.

Error rates for single-qubit gates are estimated from single-qubit randomized benchmarking on the 16-qubit register. We find an error rate per gate averaged over all 16 qubits of $0.0036$ with a standard deviation of $0.0004$. The two-qubit error rate is estimated by preparing a 16-qubit GHZ state and comparing the experimentally measured fidelity to simulated fidelities from numerical simulations accounting for errors on single-qubit gates, two-qubit gates, initialization and measurements. Averaging over multiple instances of the prepared GHZ state over the course of around 13 hours gives a mean fidelity of $0.54$ with a standard deviation of $0.04$, corresponding to an estimated two-qubit error rate $p_{2q} = 0.027$ with a standard deviation of $0.005$. Typical values for qubit initialization and measurement fault rates are $p_{\mathrm{init}} = p_{\mathrm{meas}} = 0.003$ in the device under consideration~\cite{schindler2013quantum}.

\subsection{Mid-circuit measurements}
For mid-circuit measurements only a part of the register, referred to as auxiliary qubits, is supposed to be projected into the computational basis, while a part of the register, referred to as data qubits, is ideally unaffected. To avoid projecting the data qubits, their qubit encoding is transferred to $\ket{0}_{\mathrm{data, det}}=\ket{3^2\textrm{D}_{5/2}, m_J = +1/2}$ and $\ket{1}_{\mathrm{data, det}}=\ket{3^2\textrm{D}_{5/2}, m_J = -1/2}$ by applying addressed pulses with $\theta = \pi$ ($\pi$-pulses) on the transition $\ket{4^2\textrm{S}_{1/2}, m_J = -1/2} \leftrightarrow \ket{3^2\textrm{D}_{5/2}, m_J = +1/2}$ for all data qubits. Subsequently the same detection pulse as for the final detection is applied. Recoil of scattered photons from bright auxiliary qubits heat up the ion string, which would lead to reduced gate fidelities after the mid-circuit measurement. Therefore a Doppler cooling pulse is applied. To recool the ion string close to the motional ground state an additional sideband cooling step is necessary. The data qubit encoding is transferred to $\ket{0}_{\mathrm{data, cool}}=\ket{4^2\textrm{S}_{1/2}, m_J = +1/2}$ and $\ket{1}_{\mathrm{data, cool}}=\ket{4^2\textrm{S}_{1/2}, m_J = -1/2}$ by applying $\pi$-pulses on the transitions associated to the states $\ket{4^2\textrm{S}_{1/2}, m_J = \pm 1/2}$ and $\ket{3^2\textrm{D}_{5/2}, m_J = \pm 1/2}$ as sideband cooling involves illuminating the ion string with \SI{854}{\nano\meter} light, which would otherwise destroy any information encoded in the $\ket{3^2\textrm{D}_{5/2}}$ manifold. Then the same sideband cooling pulse scheme as for state preparation is applied, apart from the fact that axial modes are not cooled and ions encoding data qubits are excluded from the set of allowed cooling ions. Cooling axial modes would require using the axial \SI{729}{\nano\meter} beam, as the addressed beam does not have overlap with the direction of motion of axial modes, and therefore would also affect data qubits. Prior to every sideband cooling pulse the respective ion is pumped to $\ket{4^2\textrm{S}_{1/2}, m_J = -1/2}$ by applying two repetitions of a $\pi$-pulse on the transition $\ket{4^2\textrm{S}_{1/2}, m_J = +1/2} \leftrightarrow \ket{3^2\textrm{D}_{5/2}, m_J = -3/2}$ followed by a \SI{5}{\micro\second} pulse of \SI{854}{\nano\meter} light. The mid-circuit measurement is finalized by repeating the pumping cycle for all auxiliary qubits that are supposed to be reused four times, and restoring the encoding of the data qubits by applying a $\pi$-pulse on the transition $\ket{4^2\textrm{S}_{1/2}, m_J = +1/2} \leftrightarrow \ket{3^2\textrm{D}_{5/2}, m_J = -1/2}$. The mid-circuit sideband cooling procedure requires around \SI{15}{\milli\second}.

\subsection{Dynamical decoupling}
\begin{figure}
    \centering
    \includegraphics[width=\linewidth]{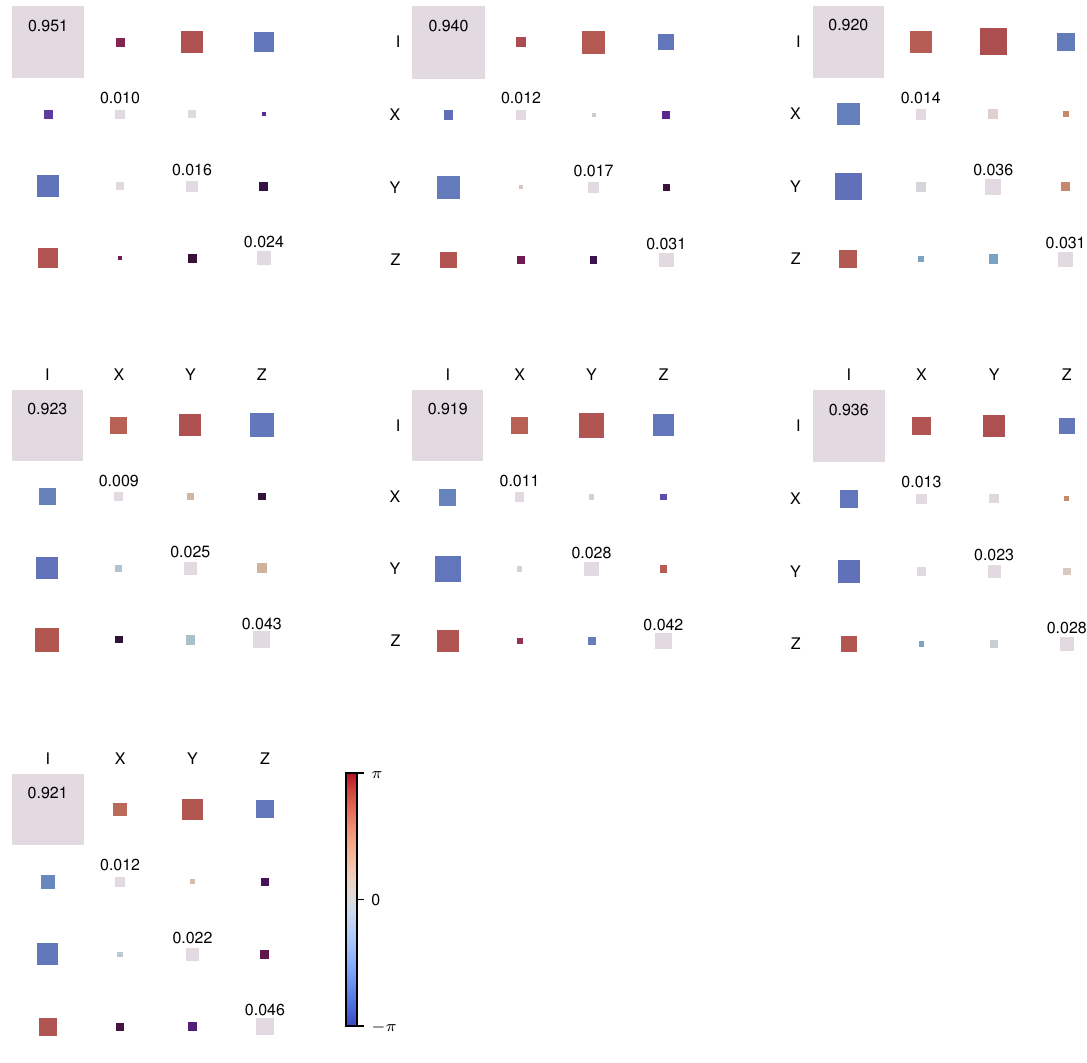}
    \caption{\justifying \textbf{Single-qubit process tomography of mid-circuit measurements.} Chi matrix representation of the process acting on data qubits during mid-circuit measurements shown for all data qubits individually. The process fidelity ranges from $0.919$ to $0.951$ with an average of $0.930$ and a standard deviation of $0.011$. The area and the color coding of the squares correspond to the absolute value and the phase of the elements of the chi matrix, respectively.}
    \label{fig:midcircuit_chi}
\end{figure}

The coherence time in the optical-qubit encoding and the ground-state encoding is on the order of \SI{50}{\milli\second} and \SI{5}{\milli\second}, respectively. Idling data qubits would thus suffer from significant dephasing during sideband cooling if no countermeasures were taken. Therefore a dynamical decoupling sequence is performed on the data qubits during the recooling procedure to preserve coherence. This decoupling is implemented with a RF antenna radiating at \SI{16.7}{\mega\hertz} acting on the entire register simultaneously which drives the transition between the two ground states, where the data qubits are encoded during sideband cooling. The antenna with a diameter of about \SI{2}{\centi\meter} is connected to resonant circuit and is mounted outside the vacuum chamber as close as possible to the ion string. Driving the resonant antenna with a power of approximately \SI{2}{\watt} allows us to implement a bit-flip in \SI{35}{\micro\second}. A decoupling pulse is applied approximately every millisecond in between cooling pulses for different motional modes. Under the application of this decoupling scheme we do not see any significant dephasing up to a waiting time of \SI{60}{\milli\second}, which indicates an effective coherence time larger than \SI{1}{\second}. The effect of a full mid-circuit measurement on data qubits is characterized via single-qubit process tomography of the data qubits using linear reconstruction. Figure~\ref{fig:experiment}C shows the chi matrix representation~\cite{nielsen2010quantum} of the process averaged over all data qubits in the Pauli basis, whereas Fig.~\ref{fig:midcircuit_chi} shows the underlying process matrices for the individual data qubits. The average fidelity is $0.930$ with a standard deviation of $0.011$. The averaged process matrix data is also used to inform the error model described in Appendix~\ref{app:noise_model}, as there are no salient differences between the individual matrices. These error probabilities are extracted from the experimental process matrix, quantifying the effect of mid-circuit measurements on data qubits, shown in Fig.~\ref{fig:experiment}C.

\section{Uncertainty estimation}\label{app:uncertainties}
The uncertainties given in this work account for statistical errors under the assumption of an underlying binomial distribution of the measurement outcomes.  We make use of the Wilson score interval~\cite{wilson1927} in order to get error intervals in the interval $\left[0, 1\right]$ even for probabilities close to $0$ or $1$. The upper (referred to as '$+$' in the formula) and lower ('$-$') bound of the interval for a probability $p$ measured with $N$ shots are given by
\begin{equation}
    p_{\pm}(p) = \frac{1}{1 + \frac{z(\alpha)^2}{N}} \left(p + \frac{z(\alpha)^2}{2N} \pm z(\alpha) \sqrt{\frac{p (1 - p)}{N} + \frac{z(\alpha)^2}{4N^2}}\right)\text{,}
\end{equation}
where $z(\alpha) = \Phi^{-1}\left(1 - \frac{\alpha}{2}\right)$ with $\Phi^{-1}$ being the quantile function of the normal distribution and $\alpha$ being the target error rate. We choose $z = 1$ which corresponds to a confidence level of $1 - \alpha \approx 0.68$.

\section{Logical fidelity}\label{app:logical_fidelity}
The figure of merit for the quality of a logical state we choose in this work is the logical fidelity, which is the probability of retrieving the correct logical state. A single logical qubit state $\rho$ is given by
\begin{align}
    \rho &=  \frac{1}{2}\left(\sigma_0 + \langle X_L \rangle \sigma_1+ \langle Y_L \rangle \sigma_2+ \langle Z_L \rangle \sigma_3\right)\text{,}
\end{align}
where $X_L$, $Y_L$ and $Z_L$ are the logical operators and $\sigma_\bullet$ are single-qubit Pauli matrices. Then the logical fidelity of $\rho$ with respect to a logical target state $\rho_t$ after performing an ideal round of QEC is given by
\begin{align}
    \mathcal{F}_t(\rho) &= \langle P_t \rangle = \mathrm{tr}\left( P_t \rho \right)\text{,}
\end{align}
with $P_t$ being the projector on the logical target state. For the logical Pauli states $|0\rangle_L$ and $|+\rangle_L$ the projectors read
\begin{equation}
    P_{|0\rangle_L} = \frac{1}{2}\left(\mathbb{1} + Z_L\right) \quad \text{and} \quad P_{|+\rangle_L} = \frac{1}{2}\left(\mathbb{1} + X_L\right) \text{,}
\end{equation}
leading to the expressions
\begin{align}
    \mathcal{F}_{|0\rangle_L} = \frac{1}{2}\left(1 + \langle Z_L \rangle\right) \quad \text{and} \quad \mathcal{F}_{|+\rangle_L} = \frac{1}{2}\left(1 + \langle X_L \rangle\right) \text{}
\end{align}
for the logical fidelities of the logical states considered in this work.

\section{Additional results}\label{app:halfcycles}
\begin{figure}
	\centering
	\includegraphics[width=0.67\linewidth]{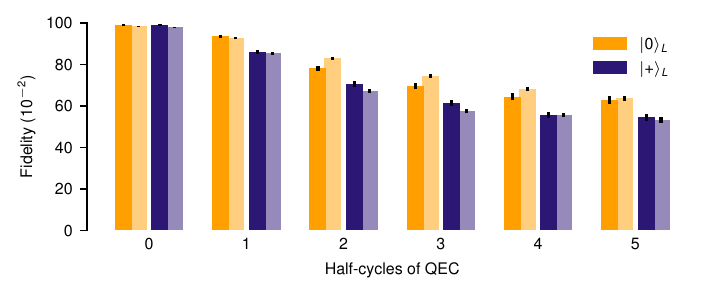}
	\caption{\justifying \textbf{Logical fidelities for half-cycles of syndrome extraction on the seven-qubit color code.} Logical fidelities obtained from Steane-type QEC for the logical input states $|0\rangle_L$ and $|+\rangle_L$. For the input states $|0\rangle_L$ and $|+\rangle_L$ only the syndrome given by the $Z$-type and $X$-type stabilizer generators, respectively, is extracted multiple times. Half-cycle $0$ corresponds to the encoding of the logical state with no extra round of QEC. The experimental and simulation results are depicted with darker and lighter shades, respectively.}
	\label{fig:half_rounds_steane}
\end{figure}

In addition to the implementation of multiple rounds of Steane QEC on the seven-qubit color code we realize repetitive readout of a single type of stabilizer generators, corresponding to executing either the first or the second half of the scheme displayed in Fig.~\ref{fig:overview_schemes_and_codes}C. The syndrome extraction is applied to a logical Pauli state sensitive to the corrections suggested by the syndrome measurement, e.g. $Z$-type stabilizer generators are measured for the input state $|0\rangle_L$ and $X$-type stabilizer generators are measured for the input state $|+\rangle_L$. We refer to one readout as a half-cycle of QEC. We implement up to five half-cycles of QEC with the corresponding logical fidelities being shown in Fig.~\ref{fig:half_rounds_steane}. Again we see good agreement of experiments with data from numerical simulations.

\clearpage

\end{document}